\documentclass{aa}
\usepackage{amsmath,graphicx,amssymb}
\usepackage{txfonts}
\usepackage{natbib}
\bibpunct{(}{)}{;}{a}{}{,}

\newcommand\hvezda{\object{CU~Vir}}
\newcommand{\zav}[1]{\left(#1\right)}
\newcommand{\hzav}[1]{\left[#1\right]}
\newcommand\intvidpo{\!\!\int\limits_{\begin{array}{c}\text{\scriptsize
visible}\\[-2mm]\text{\scriptsize surface}\end{array}}\!\!}

\newlength\staretab

\makeatletter
\def\sgn{\mathop{\operator@font sgn}\nolimits}
\makeatother

\begin{document}

\title{Modelling of the ultraviolet and visual SED variability in the hot magnetic Ap star CU Vir}

\author{J.~Krti\v{c}ka\inst{1} \and Z.~Mikul\'a\v sek\inst{1,2}
        \and T. L\"uftinger\inst{3} \and D. Shulyak\inst{4}
        \and J.~Zverko\inst{5} \and J.~\v Zi\v z\v novsk\'y\inst{6}
        \and N.~A.~Sokolov\inst{7}}

\offprints{J.~Krti\v{c}ka,\\  \email{krticka@physics.muni.cz}}

\institute{Department of Theoretical Physics and Astrophysics,
           Masaryk University, Kotl\'a\v rsk\' a 2, CZ-611\,37 Brno, Czech Republic
            \and
            Observatory and Planetarium of J. Palisa, V\v SB -- Technical
            University, Ostrava, Czech Republic
            \and
            Institut f\"ur Astronomie, Universit\"at Wien,
            T\"urkenschanzstra\ss e 17, 1180 Wien, Austria
            \and
            Institute of Astrophysics, Georg-August-University,
            Friedrich-Hund-Platz 1, D-37077 G\"ottingen, Germany
            \and
            Tatransk\'{a} Lomnica 133, SK-05960 Slovak Republic
            \and
            Astronomical Institute, Slovak Academy of Sciences,
            Tatransk\'{a} Lomnica, SK-05960 Slovak Republic
            \and
            Central Astronomical Observatory at Pulkovo, St. Petersburg
            196140, Russia
            }

\date{Received}

\abstract
{The spectral energy distribution (SED) in chemically peculiar
stars may be significantly affected by their abundance anomalies. The
observed SED variations are usually assumed to be a result of
inhomogeneous surface distribution of chemical elements, flux
redistribution and stellar rotation. However, the direct evidence for
this is still only scarce.}
{We aim to identify the processes that determine the SED and its variability in the UV and
visual spectral domains of the helium-weak star CU Vir.}
{We used the TLUSTY model atmospheres calculated for the appropriate surface
chemical composition to obtain the emergent flux and predict the rotationally
modulated flux variability of the star.} {We show that most of the light
variations in the $vby$ filters of the Str\"omgren photometric system
are a result of the uneven surface distribution of
silicon, chromium, and iron. Our models are only able to explain
a part of the variability in the $u$ filter, however. The observed UV flux
distribution is very well reproduced, and the models are able to
explain most of the observed features in the UV light curve, except for the
region $2000-2500\,$\AA, where the amplitude of the observed light
variations is higher than predicted. The variability observed in the
visible is merely a faint gleam of that in the UV. While the amplitude
of the light curves reaches only several hundredths of magnitude in
the visual domain, it reaches about 1\,mag in the UV.} {The visual and
UV light variability of CU Vir is caused by the flux redistribution
from the far UV to near UV and visible regions, inhomogeneous
distribution of the elements and stellar rotation. Bound-free transitions
of silicon and bound-bound transitions of iron and chromium contribute the most 
to the flux redistribution. This mechanism can
explain most of the rotationally modulated light variations in the
filters centred on the Paschen continuum and on the UV continuum
of the star CU Vir. However, another mechanism(s) has
to be invoked to fully explain the observed light variations in the $u$ filter
and in the region $2000-2500\,$\AA.}


\titlerunning{Modelling of the ultraviolet and visual SED variability in hot
magnetic Ap star CU Vir}
\authorrunning{J.~Krti\v{c}ka et al.}
\maketitle

\section{Introduction}

Chemically peculiar (CP) stars are among the most enigmatic objects of
the upper part of the main-sequence. The processes of radiative diffusion and
gravitational settling in the atmospheres of these stars cause pronounced
deviations  from the solar value in the chemical composition (\citealt{vaupreh},
\citealt{mpoprad}). Apart from the chemical peculiarity, many of CP stars
show variations in the magnetic field as well as in their spectra and light.
These variations are usually
strictly periodical and modulated by the rotation of the star. The uneven
distribution of the surface magnetic field (with dipole component dominating in
most cases) is one of the factors that cause the uneven distribution of
chemical elements and, consequently, also the periodic spectrum variability. The
uneven distribution of chemical elements is suspected to be the origin of the
light variability, but this connection is still not very well understood.

The line blanketing caused by numerous lines of overabundant elements (mainly the
iron-peak ones) and the flux redistribution induced by these lines has been
suspected to significantly affect the spectral energy distribution (SED) and to be the prime source of the light
variability \citep[e.g.,][]{molnar,kodcar}. Bound-free transitions
\citep{peter,lanko} were also expected to play some role. Surface temperature
differences or variable temperature gradients \citep{biltep,steptep} were 
on the list of possible causes of the SED variability as well. The presence of a magnetic
field may affect the SED and its variability, provided the field is sufficiently strong
\citep{malablaj}. Finally, hot CP stars may have circumstellar shells fed by the
stellar wind, causing variability through the light absorption
\citep{labor,nakaji,smigro,towog}.

Realistic SED simulations of CP stars were not possible until
the techniques of Doppler mapping and model atmosphere calculations were
considerably developed. The Doppler mapping enables one to precisely map the
distribution of individual elements on the surface of rotating stars
\citep[e.g.,][]{ryze,choch,piskoc,lroap,bohacen}. Detailed model atmospheres
\citep[e.g.,][]{bstar2006} enable one to precisely predict the fluxes from surface
elements with a peculiar chemical composition, taking into account realistic
bound-free \citep{topt} and bound-bound \citep{kur22} transitions.

Based on precise model atmospheres and detailed surface maps it was possible to
follow earlier attempts of light curve modelling \citep{krivo,rytep} and to
successfully simulate the light curves of several CP stars. \citet{myhd37776}
showed that the light variations of \object{HD 37776}
($T_\text{eff}=22\,000\,\text{K}$) can be explained to be a result of inhomogeneous
surface distribution of helium and silicon. \citet{myhr7224} showed that most of
the observed light variations in \object{HR 7224}
($T_\text{eff}=14\,500\,\text{K}$) are caused by inhomogeneous surface distribution of
silicon and iron. For the cooler star \object{$\varepsilon$ UMa}
($T_\text{eff}=9\,000\,\text{K}$) \citet{seuma} showed that chromium can also 
contribute significantly to the light variability. Moreover, modern atmosphere
models are able to explain the observed SED in detail \citep[e.g.,][]{zeminy},
pointing to the importance of rare-earth elements in cooler CP stars.

For our present study we selected one of the most enigmatic CP stars, \hvezda\
(HR~5313, HD~124224).
The light variability of \hvezda\ has been known for more than half a century
\citep{laurel}. \hvezda\ belongs to a rare group of CP stars that show period
changes \citep{pyperper,pypadel,trigi08,trigi11,mikvar}, theoretically studied
by \citet{stepbrz}. Moreover, \hvezda\ is a source of variable radio emission,
resembling a radio lighthouse of pulsars \citep{trigilio,kellett}.

All these observations make \hvezda\ one of the most appealing targets for
theoretical studies. We studied the nature of the SED variations of this star
using Doppler abundance maps of \citet{kus}.

\section{Simulation of the SED variability}

\subsection{Stellar parameters}
\label{stelpa}

The stellar parameters of \hvezda\ and abundance maps of helium, silicon,
chromium, and iron adopted from \citet{kus} are given in Table~\ref{hvezda}.
Note that there is also a magnesium abundance map available in \citet{kus},
but because of the low maximum magnesium abundance derived we did not include its
inhomogeneous surface distribution. The abundances in the maps are expressed as
$\log\zav{N_\text{el}/N_\text{tot}}$, but we used abundances relative
to hydrogen, i.e., $\varepsilon_\text{el}=\log\zav{N_\text{el}/N_\text{H}}$.

\begin{table}[t]
\caption{\hvezda\ parameters from spectroscopy \citep{kus}.}
\label{hvezda}
\begin{center}
\begin{tabular}{lc}
\hline
Effective temperature ${{T}_\mathrm{eff}}$ & ${13\,000}$\,K \\
Surface gravity ${\log g}$ (cgs) & ${4.0}$ \\
Inclination ${i}$ & ${30^\circ}$ \\
Rotational velocity projection $v_\text{rot} \sin i$ & $160\,\text{km}\,\text{s}^{-1}$\\
Helium abundance&$-3.2<\varepsilon_\text{He}<-1$ \\
Silicon abundance& $-4.6<\varepsilon_\text{Si}<-2.3$ \\
Chromium abundance& $-6.6<\varepsilon_\text{Cr}<-4.4$ \\
Iron abundance&$-5.5<\varepsilon_\text{Fe}<-3.5$  \\
\hline
\end{tabular}
\end{center}
\end{table}

The calculation of the rotational phases for individual observations
is not a straightforward task, because the instant rotational
period $P(t)$ of the surface layers of \hvezda\ is changing. Accordingly,
we applied the new ephemeris of \citet{mikvar},
where the phase function $\vartheta(t)$ is approximated by
the fourth-order polynomial of time:
\begin{align}\label{cuteta}
\vartheta(t)&\cong  \vartheta_0-\frac{A}{P_0} \zav
{\textstyle{\frac{3}{2}}\mathit{\Theta}^2-\mathit {\Theta}^4};\
\vartheta_0=\frac{t-M_0}{P_0},\
\mathit{\Theta}=\frac{t-T_0}{\mathit{\Pi}},\\
P(t)&=  1/\dot{\vartheta}\doteq
P_0\hzav{1+A/\mathit{\Pi}\zav{3\mathit{\Theta}-4\,\mathit{\Theta}^3}},
\end{align}
where $P(t)$ is the instant period at the time $t$,
$\vartheta_0$ is the phase function for a linear ephemeris with the
origin at $M_0 \equiv 2\,446\,730.4447$ and the basic period $P_0$.
\citet{mikvar} found that $P_0 = 0\fd52069415(8)$,
$A=0\fd5643(29)$, $\mathit{\Pi} = 13260(70)$\,d, and $T_0 =
2\,446\,636(24)$. The formula accounts for the period variability
observed in \hvezda\ and enables us to determine the rotational phase
with an accuracy better than 0.002 $P$.

The phase shift between this ephemeris
$\phi=\mathrm{frac}(\vartheta(t))$ (fractional part of
$\vartheta(t)$) and that used by \citet{kus} $\phi_\text{Kus}$ in the
time of their spectral observations was
$\Delta\phi=\phi-\phi_\text{Kus}=0.52045$, and the origin of the phase
function was determined to be at HJD 2\,446\,730.4447.

\subsection{Model atmospheres and synthetic spectra}

We used the code TLUSTY for the model atmosphere calculations
\citep{tlusty,hublaj,hublad,lahub}. Although the code enables us to
calculate NLTE models, we confined ourselves to the LTE plane-parallel
models, because we expected the NLTE effects to be marginal for the
light variability. The atomic data \citep[taken from][]{bstar2006} were
selected to be appropriate for B type stars, the atomic
data for silicon  in particular are based on \citet{mendo}, \citet{maslo93}, and
Taylor (2011), in preparation\footnote{Note that the opacity caused by
autoionisation is included via bound-free cross section (as default
in TLUSTY).}; for iron on \citet{kur22}, \citet{nah96}, \citet{nah97},
\citet{bau97}, and \citet{bau96}, and for other elements on
\citet{top1}, \citet{topf}, \citet{toptul}, \citet{topp},
\citet{toph}, and \citet{napra}. We prepared our own ionic models for
chromium (\ion{Cr}{ii}--\ion{Cr}{vi}) using data taken from Kurucz
(2009)\footnote{http://kurucz.harvard.edu}.

We assumed fixed values of the effective temperature and surface gravity
(according to Table~\ref{hvezda}) and adopted a generic value of the
microturbulent velocity $v_\text{turb}=2\,\text{km}\,\text{s}^{-1}$. The
abundance of helium, silicon, chromium, and iron differed in individual models as
explained below. We used the solar abundance of other elements \citep{asgres}.

For the calculation of synthetic spectra we used the SYNSPEC code. The synthetic
spectra were calculated for the same parameters (effective temperature, surface
gravity, and chemical composition) as the model atmospheres. We also took into
account the same transitions as for the model atmosphere calculations. To this
end we included the same chromium and iron lines as we used for the model atmosphere
calculation in our SYNSPEC line list. This is not particularly important in the
visible, but in the ultraviolet (UV) numerous lines for which only the
theoretical data are available significantly influence the spectral energy
distribution. Additionally, we included the lines of all elements with the
atomic number $Z\leq30$, that were not accounted fot the model atmosphere calculation. We
computed angle-dependent intensities for $20$ equidistantly spaced values of
$\mu=\cos\theta$, where $\theta$ is the angle between the normal to the surface
and the line of sight.

The model atmospheres and the angle-dependent intensities
$I(\lambda,\theta,\varepsilon_\text{He},\varepsilon_\text{Si},\varepsilon_\text{Cr},\varepsilon_\text{Fe})$
mentioned above were calculated for a four-parametric grid of helium, silicon,
chromium, and iron abundances (see Table~\ref{esit}). For silicon this grid
fully covers the range of silicon abundances in the map of \citet{kus}, but
for helium, chromium, and iron the lowest abundances detected by \citet{kus} are
omitted from the grid. Our test showed that this restriction of the grid does
not influence the predicted light curves. The generation of the complete grid
would require the calculation of 300 model atmospheres and synthetic spectra.
Because not all abundance combinations are required for the interpolation of the
\citet{kus} maps, we calculated only those models that were necessary. This
helped us to reduce the number of calculated models by half.

\begin{table}[t]
\caption{Individual abundances $\varepsilon_\text{He}$, $\varepsilon_\text{Si}$,
$\varepsilon_\text{Cr}$, and $\varepsilon_\text{Fe}$ of the model grid}
\label{esit}
\begin{center}
\begin{tabular}{lrrrrrrr}
\hline
He& $-2.0$ &$-1.0$\\
Si& $-4.75$& $-4.25$& $-3.75$& $-3.25$& $-2.75$ & $-2.25$\\
Cr& $-6.4$& $-5.9$& $-5.4$& $-4.9$& $-4.4$\\
Fe & $-5.4$& $-4.9$& $-4.4$& $-3.9$& $-3.4$ \\
\hline
\end{tabular}
\end{center}
\end{table}

\subsection{Phase-dependent flux distribution}
\label{vypocet}

The radiative flux in a colour $c$ at the distance $D$ from the star with
radius $R_*$ is \citep{mihalas}
\begin{equation}
\label{vyptok}
f_c=\zav{\frac{R_*}{D}}^2\intvidpo I_c(\theta,\Omega)\cos\theta\,\text{d}\Omega,
\end{equation}
where the intensity $I_c(\theta,\Omega)$ at each surface point with
spherical coordinates $\Omega$ is obtained by means of
interpolation between the intensities
$I_c(\theta,\varepsilon_\text{He},\varepsilon_\text{Si},\varepsilon_\text{Cr},\varepsilon_\text{Fe})$
calculated from the grid of synthetic spectra (see Table~\ref{esit}) as
\begin{equation}
\label{barint}
I_c(\theta,\varepsilon_\text{He},\varepsilon_\text{Si},\varepsilon_\text{Cr},\varepsilon_\text{Fe})=
\int_0^{\infty}\Phi_c(\lambda) \,
I(\lambda,\theta,\varepsilon_\text{He},\varepsilon_\text{Si},\varepsilon_\text{Cr},\varepsilon_\text{Fe})\, \text{d}\lambda.
\end{equation}
The transmissivity function $\Phi_c(\lambda)$ of a given filter $c$
of the Str\"omgren photometric system is
approximated for simplicity by a Gauss function
\citep[see][for details]{myhr7224}.

The magnitude difference is defined as
\begin{equation}
\label{velik}
\Delta m_{c}=-2.5\,\log\,\zav{\frac{{f_c}}{f_c^\mathrm{ref}}},
\end{equation}
where $f_c$ is calculated from Eq.~\ref{vyptok} and
${f_c^\mathrm{ref}}$ is the reference flux obtained under the
condition that the mean magnitude difference over the rotational period is zero.

\section{Influence of the abundance on the emergent flux}
\label{kaptoky}

\begin{figure}[tp]
\centering \resizebox{\hsize}{!}{\includegraphics{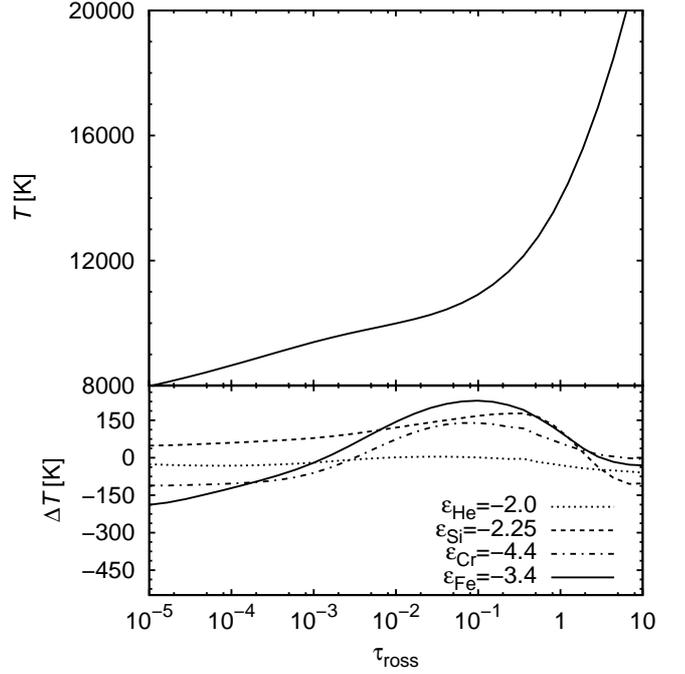}}
\caption{{\em Upper plot:} The dependence of temperature on the Rosseland
optical depth $\tau_\text{ross}$ in
the reference model atmosphere with $\varepsilon_\text{He}=-1.0$,
$\varepsilon_\text{Si}=-3.75$, $\varepsilon_\text{Cr}=-5.9$, and
$\varepsilon_\text{Fe}=-4.4$. {\em Lower plot}: The temperature in the
model atmospheres with modified abundance of individual elements minus the
temperature in the reference model atmosphere.}
\label{tep}
\end{figure}

Individual elements modify the temperature distribution of model
atmospheres by their bound-free and bound-bound transitions. This can
be seen in Fig.~\ref{tep}, where we compare the temperature distribution
of model atmospheres for typical abundances found on the surface of
\hvezda. The bound-free (caused by ionisation of helium and silicon) and
bound-bound transitions (line transition of chromium and iron) absorb
the stellar radiation, consequently the temperature in the continuum-forming
region ($\tau_\text{ross}\approx0.1-1$) increases with
increasing abundance of these elements. For silicon  and iron the
influence of abundance on the temperature is stronger, for chromium
the influence is weaker, while for typical helium abundances found on
the surface of \hvezda\ the changes of temperature are only marginal.

\begin{figure*}[tp]
\centering \resizebox{0.75\hsize}{!}{\includegraphics{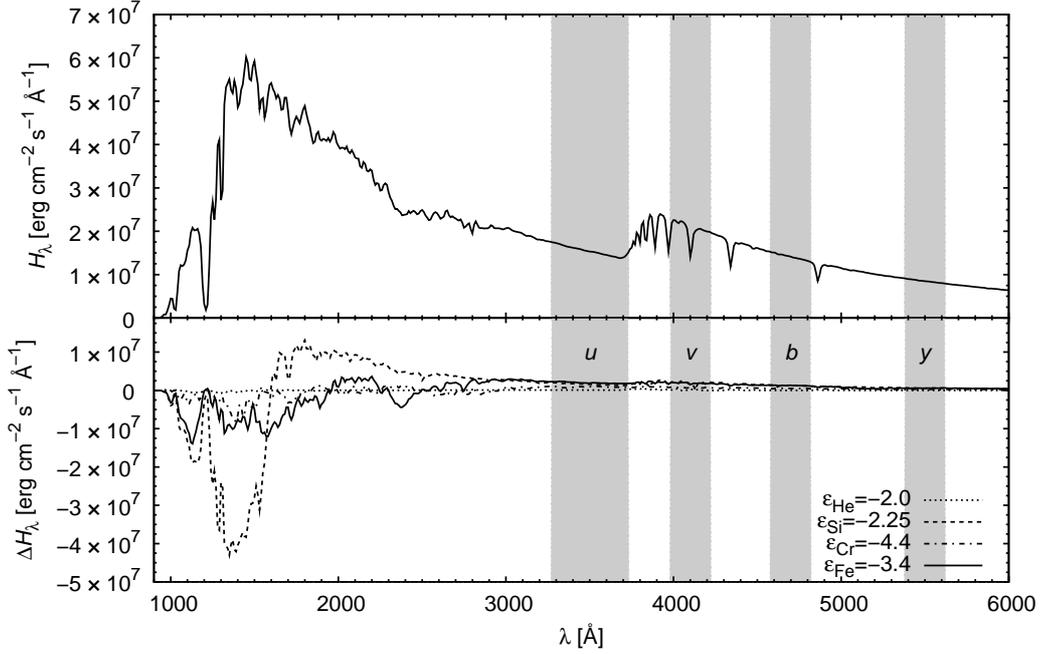}}
\caption{{\em Upper plot:} The emergent flux from a reference model
atmosphere with $\varepsilon_\text{He}=-1.0$, $\varepsilon_\text{Si}=-3.75$,
$\varepsilon_\text{Cr}=-5.9$, and $\varepsilon_\text{Fe}=-4.4$. {\em Lower
plot}: The emergent flux from the model atmospheres with modified abundance of
individual elements minus the flux from a reference model. All fluxes were
smoothed by a Gaussian filter with a dispersion of $10\,\AA$ to show the changes
in continuum, which are important for SED variability. The passbands of
the $uvby$ photometric system are also shown in the graph (grey
areas).}
\label{prvtoky}
\end{figure*}

In atmospheres with overabundant helium, silicon, chromium or iron the enhanced
opacity leads to the redistribution of the flux from the short-wavelength part of
the spectrum to the longer wavelengths of the UV spectrum, and also to the
visible spectral regions (see Fig.~\ref{prvtoky}). Consequently, the
overabundant spots are bright in the $uvby$ colours, and are dark in
far-ultraviolet bands. As already found by \citet{myhd37776}, helium can affect the
flux distribution only if it significantly dominates over hydrogen, i.e. for
$\varepsilon_\text{He}>0.5$. Consequently, for model atmospheres with
underabundant helium the flux variations are only marginal. Note also that
the flux variations caused by silicon are most pronounced in the far-UV region with
$\lambda<1600\,$\AA.

\begin{figure}[t]
\centering
\resizebox{0.9\hsize}{!}{\includegraphics{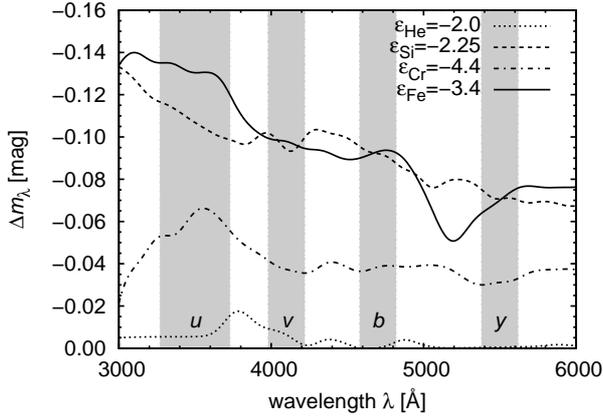}}
\caption{Magnitude difference $\Delta m_\lambda$ between the emergent fluxes
calculated  with an enhanced abundance of individual elements and the reference flux
$H_\lambda^\text{ref}$ (see Eq.~\ref{tokmagroz}).
The fluxes were smoothed with a Gaussian filter with a dispersion
of $100\,$\AA.}
\label{magtoky}
\end{figure}

These flux changes can be detected as a change in the apparent magnitude. To
demonstrate this, we plot (Fig.~\ref{magtoky}) the relative magnitude difference
defined as
\begin{equation}
\label{tokmagroz}
\Delta m_\lambda=-2.5\log\zav{
\frac{H_\lambda(\varepsilon_\text{He},\varepsilon_\text{Si},
\varepsilon_\text{Cr},\varepsilon_\text{Fe})} {H_\lambda^\text{ref}}},
\end{equation}
against wavelength. Here $H_\lambda^\text{ref}$ is the reference flux calculated
for slightly overabundant chemical composition (with
$\varepsilon_\text{He}=-1.0$, $\varepsilon_\text{Si}=-3.75$,
$\varepsilon_\text{Cr}=-5.9$, and $\varepsilon_\text{Fe}=-4.4$). As can be seen
in Fig.~\ref{magtoky}, the absolute value of the relative magnitude difference
decreases with increasing wavelength. However, the behaviour of the flux
calculated for modified helium is different. The maxima of the relative
brightness at the positions of the hydrogen lines (especially close to the Balmer
jump) remained undetected in the previous analysis. These maxima are caused by
the strengthening of the Lorentz wings of the hydrogen lines owing to the higher
density in the line-forming region in the models with higher helium abundance.
It has not escaped our attention that the minimum of the relative magnitude
difference at about $5200\,$\AA\ caused by an accumulation of iron lines can be
connected with the well-known flux depression at these wavelengths \citep[as
discussed already by, e.g.,][]{preslo,myhr7224}.

\section{Predicted light variations}

\begin{figure}[t]
\centering
\resizebox{0.9\hsize}{!}{\includegraphics{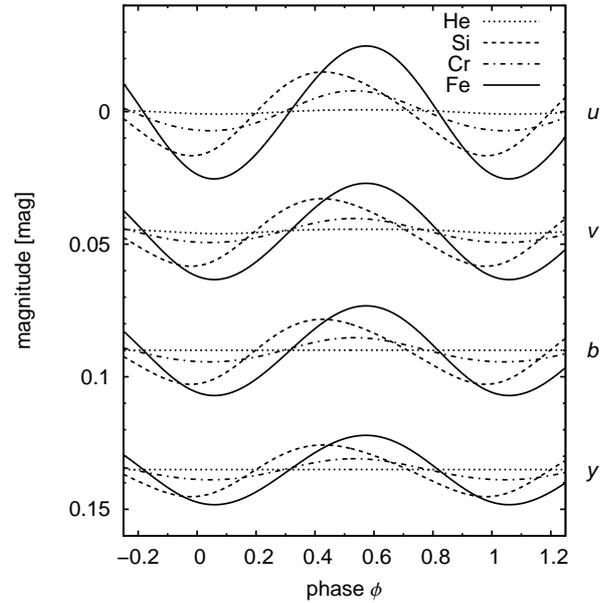}}
\caption{Predicted light variations of \hvezda\ in the Str\"omgren photometric
system calculated using abundance maps of one element only. The abundance of
other elements was fixed. Light curves in individual filters were
vertically
shifted to
better demonstrate the light variability.}
\label{prv_hvvel}
\end{figure}

\begin{figure}[t]
\centering
\resizebox{0.9\hsize}{!}{\includegraphics{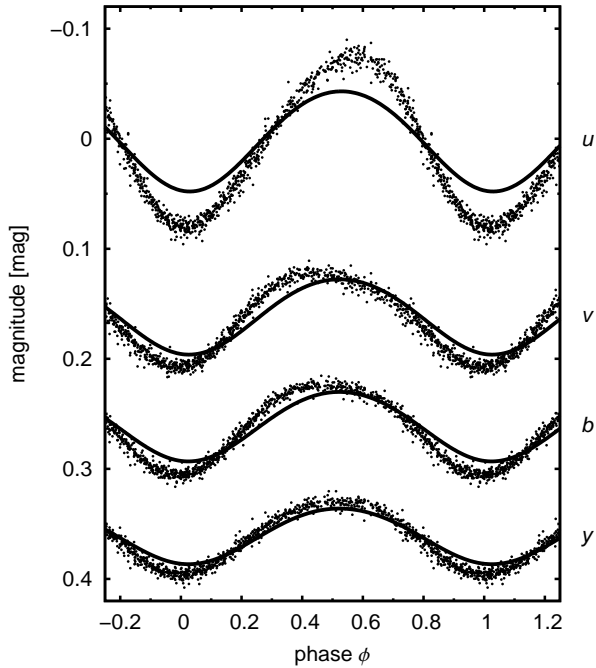}}
\caption{Predicted light variations of \hvezda\ (solid lines) computed taking
into account the helium, silicon, chromium, and iron surface abundance distributions
derived by \citet{kus}. The observed light variations (dots) are taken from
\citet{pyperper}.
The light curves in individual filters were
shifted vertically to better demonstrate the light
variability.}
\label{cuvir_hvvel}
\end{figure}

Predicted light curves are calculated from the surface abundance maps
derived by \citet{kus} and from the emergent fluxes computed with the
SYNSPEC code, applying Eq.~\ref{velik} for individual rotational phases.

To study the influence of individual elements separately, we first calculated
the light variations with the abundance map of one element only
(Fig.~\ref{prv_hvvel}), assuming a fixed abundance of other elements
($\varepsilon_\text{He}=-1.0$, $\varepsilon_\text{Si}=-3.75$,
$\varepsilon_\text{Cr}=-5.9$, $\varepsilon_\text{Fe}=-4.4$). From
Fig.~\ref{prv_hvvel} it follows that iron, silicon, and chromium contribute most
to the light variations, while the contribution of helium is only marginal. This
is because of the large overabundance of these elements in the spots and by their
large abundance variations on the stellar surface. The amplitude of the light
variations increases with decreasing wavelength, as can be expected from the
plot of the magnitude difference $\Delta m_\lambda$ in Fig.~\ref{magtoky}.
Because the overabundant regions are brighter in the $uvby$ colours, the
predicted light variations reflect the equivalent width variations
\citep[Fig.~1]{kus}. The light maximum occurs at the same phase at
which the equivalent widths of a given element are the largest. Because this
happens at slightly different phases for individual elements, the light curves
in Fig.~\ref{prv_hvvel} are slightly shifted.

\begin{figure}[t]
\centering
\resizebox{0.9\hsize}{!}{\includegraphics{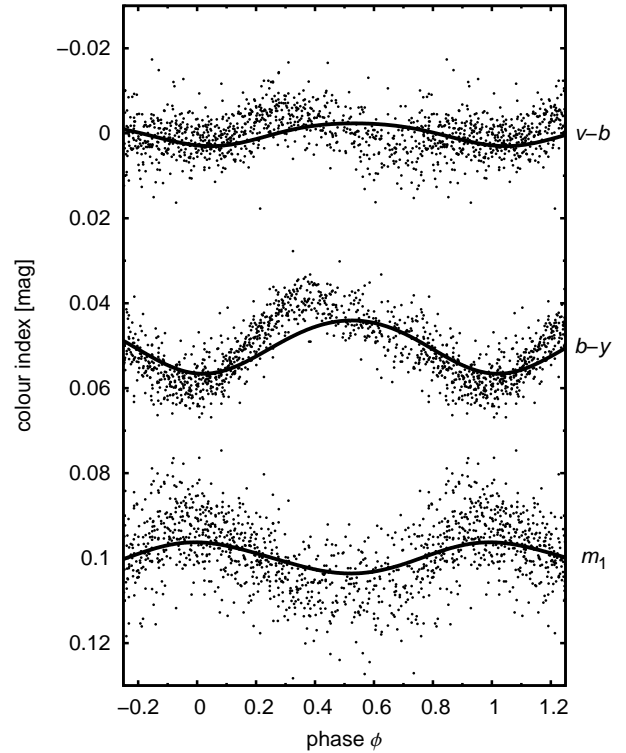}}
\caption{Predicted variations of colour indices (solid lines) calculated
from the helium, silicon, chromium, and iron surface abundance maps compared
with the observations. Observed light variations (dots) are taken from
\citet{pyperper}. Light curves in individual filters were
vertically
shifted to
better demonstrate the light variability.}
\label{cuvir_uvby}
\end{figure}

Taking into account the surface distribution of helium, silicon, chromium, and
iron in the calculation of the light curves (Fig.~\ref{cuvir_hvvel}), we
obtained
a good agreement between the observed and predicted light curves in the $v$, $b$
and $y$ bands of the Str\"omgren photometric system. On the other hand, our
models are able to explain only about half of the amplitude in the $u$ filter.
The disagreement between the predicted and observed light curves is mostly
apparent around phase $\phi=0.6$. Note also that a similar disagreement
visible in $u$ can be also found in other filters, but to a much smaller extent.
These differences clearly point to an existence of some additional, unknown
mechanism working especially in the violet band that still needs to be
investigated (Fig.~\ref{cuvir_hvvel}, and see also Sect.~\ref{kecame}). The
discrepancies between the predicted and observed light curves increase when
comparing the predicted and observed colour indices (see Fig.~\ref{cuvir_uvby}).
While  the $(b-y)$ data agree reasonably well, the $(v-b)$ curves are mutually
shifted, and the predicted metallic index $m_{\text{1}}=(v-b)-(b-y)$ shows a
significantly lower amplitude than the observed one.

\begin{figure*}[t]
\centering
\resizebox{0.6\hsize}{!}{\includegraphics{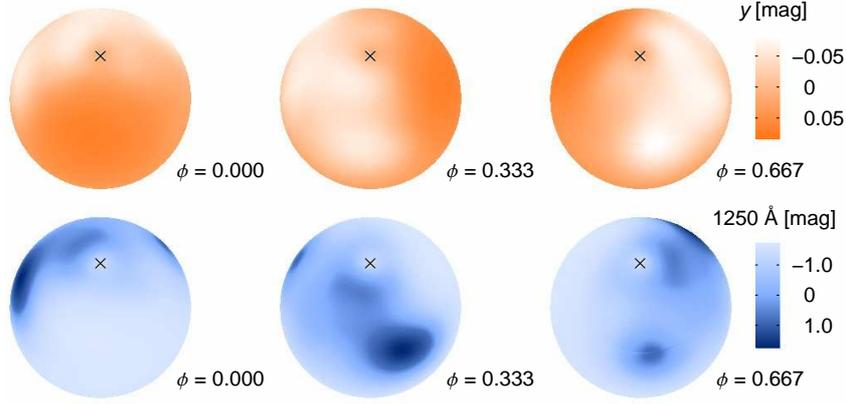}}
\caption{Emergent intensity from individual surface elements of \hvezda\ at
various rotational phases. Upper panel: visible $y$ band. Lower panel: UV band
centred at $1250\,$\AA. Both for $\mu=1$.}
\label{cuvir_povrch}
\end{figure*}

The inhomogeneous surface distribution of individual elements causes
bright spots on the stellar surface. The spots, whose surface
distribution can be derived using abundance maps and model atmospheres (see
Fig.~\ref{cuvir_povrch}), cause the light variability.

\section{Ultraviolet variations}
\label{uv}

We have shown that the light variability of \hvezda\ is caused by the
redistribution of flux from the far UV to the near UV and visible regions.
Consequently, the light variability in the far UV region should be in antiphase
with the visual one. This behaviour was indeed found in a detailed analysis of
IUE observations of \hvezda\ by \citet{sokolpan}.

To test these predictions quantitatively as well, we extracted IUE observations of
\hvezda\ from the INES database \citep[see Table \ref{iuetab}]{ines} using the
SPLAT package \citep[see also \citealt{pitr}]{splat}. Here we used low-dispersion
large aperture spectra in the domains 1250--1900~\AA\ (SWP camera) and
2000--3000~\AA\ (LWR camera).

\subsection{Narrow-band UV variations}

As a first comparison of the UV fluxes we concentrated on narrow-band
variations. For this purpose we smoothed the observed and predicted fluxes
with a Gaussian filter with a dispersion of $10\,$\AA.

\begin{figure*}[t]
\centering \resizebox{0.75\hsize}{!}{\includegraphics{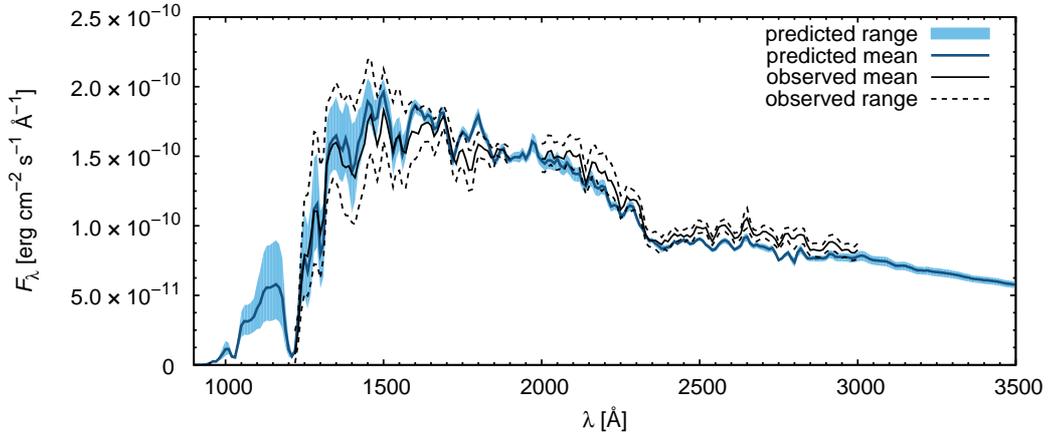}}
\caption{Comparison of the predicted flux (mean and its variation, blue)
with corresponding observed quantities (black). Both predicted and observed
fluxes (IUE) were smoothed by a Gaussian filter with a dispersion of $10\,$\AA.}
\label{ptok}
\end{figure*}

\begin{figure*}[t]
\centering \resizebox{0.49\hsize}{!}{\includegraphics{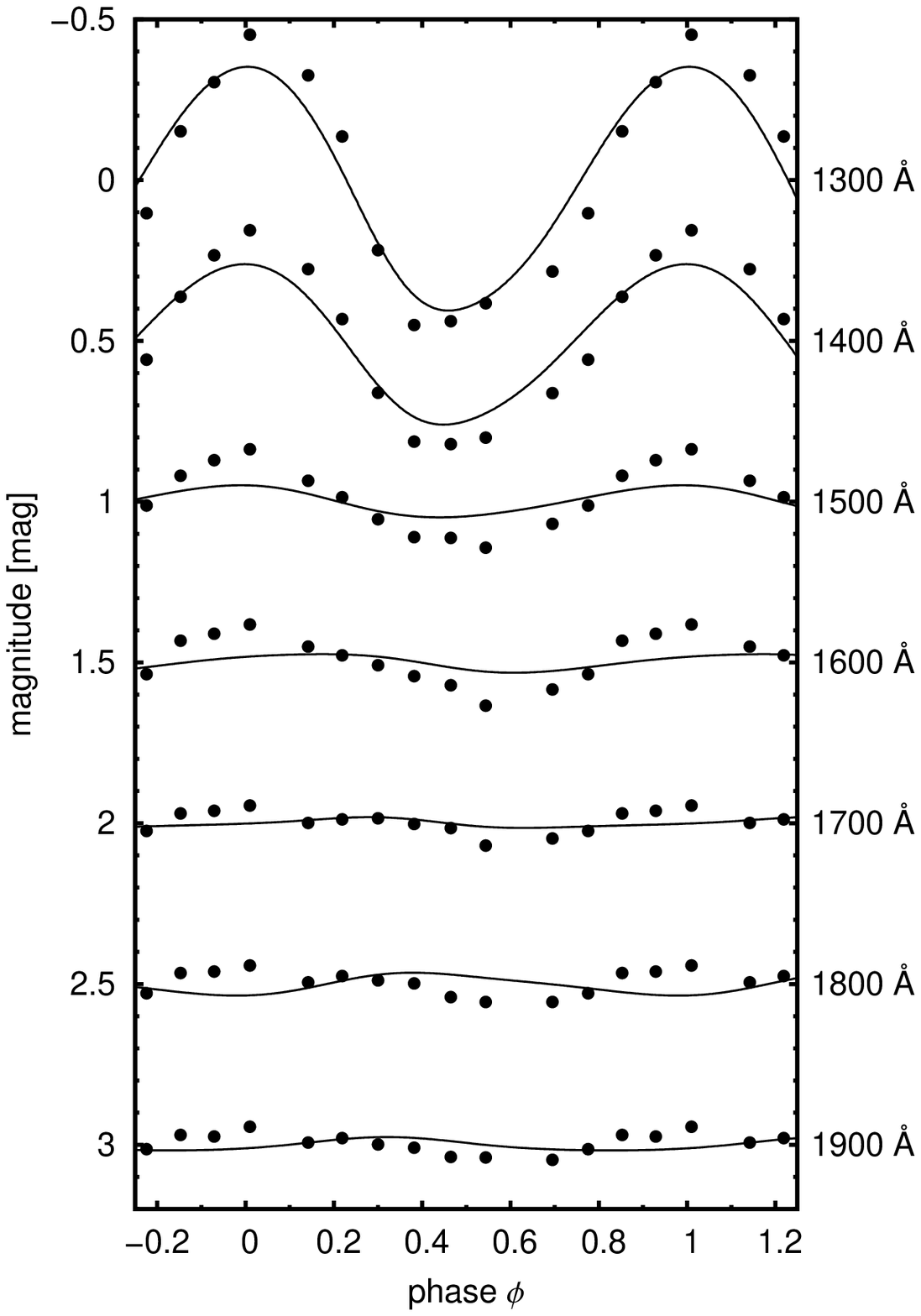}}
\centering \resizebox{0.49\hsize}{!}{\includegraphics{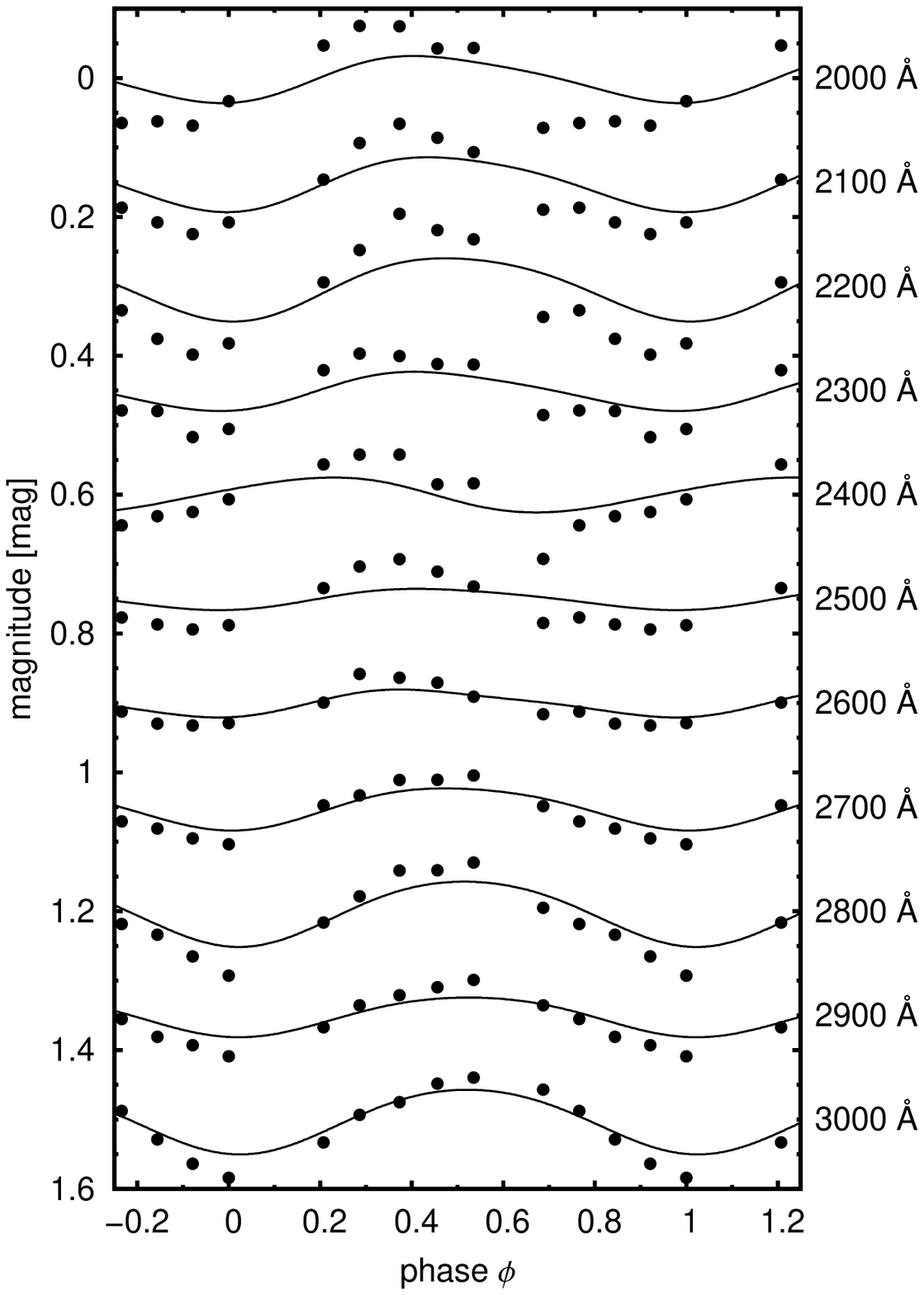}}
\caption{Comparison of the predicted (solid line) and observed (dots) UV
light variations for different wavelengths.
Curves for individual wavelengths were vertically shifted to
better demonstrate the variability.
}
\label{fuvnuv}
\end{figure*}

The resulting predicted and observed UV fluxes are given in
Fig.~\ref{ptok}. To avoid possible problems with absolute IUE
calibration, we normalised the predicted fluxes by a multiplicative
factor, which yields the best match between observations and prediction
in Fig.~\ref{ptok}. The factor was kept fixed for all wavelengths in
all subsequent calculations. Generally, 
the mean predicted and observed fluxes agree well, with some minor
differences. The mean observed flux is slightly lower than the
predicted one in the region $1600-1850\,$\AA, whereas it is slightly
higher in $2100-3000\,$\AA.  The amplitude of the observed flux
variations agrees well with the observed one in the regions
$1250-1450\,$\AA\ and $2550-3000\,$\AA. In the remaining regions the
predicted amplitude is lower than the observed one.

As can be seen from Fig.~\ref{fuvnuv}, 
the predicted and observed flux variations in the selected wavelengths agree very well. Our
simulations are able to explain most of the observed features in the UV light
curve. The amplitude is highest in the far-UV region $1250-1400\,$\AA,
reaching nearly 1\,mag at $1250\,$\AA. The light variations in this region are
mainly caused by silicon (c.f., Fig.~\ref{prvtoky}, and \citealt{sokold,sokolja}).
The overall agreement between the predicted and observed light curves in this
region indicates that silicon abundances are well mapped.

The amplitude of the light variations is very low in the region
between $1600-1900\,\AA$. The silicon-rich patches are dark in this
wavelength region, whereas the iron-rich ones are bright, causing a near
cancelation of any light variability in common. Note, however, that a
fine structure of observed variations is not completely reproduced by
the models, indicating either that the model atmospheres need to be improved or,
which is even more likely, that the light variability has other sources.

Our models are able to nicely reproduce the observed \citep{sokolpan} antiphase
variations in the far-UV on one side and the near-UV and visible region on the other
side (see also Fig.~\ref{cuvir_povrch}). However, as was already clear from
Fig.~\ref{ptok}, the observed light amplitude is higher than the predicted one
in the region $2000-2500\,$\AA. This disagreement together with the difference
in the $u$ light curves (see Fig.~\ref{cuvir_hvvel}) indicates a presence of
an additional now unidentified source of light variability. This might be
connected with a chemical element whose abundance was not mapped by \citet{kus}.
This element, together with silicon, might contribute to the light variability
in this region.

Interestingly, despite the disagreement of the observed and predicted light
variations in the $u$ colour of the Str\"omgren photometric system
(Fig.~\ref{cuvir_hvvel}), our models are able to nicely reproduce most of the
light variations in the region $2550-3000\,$\AA\ (Fig.~\ref{fuvnuv}). The
amplitudes of the observed light curves in this region and their shapes are
a result of the inhomogeneous surface distribution of silicon,
iron, and chromium \citep[see also][]{sokolja}. Even more subtle effects, like
the mutual shift of the light maxima at wavelengths $2550\,$\AA\ and
$2800\,$\AA, can be explained by our models.

\begin{figure*}[t]
\centering \resizebox{0.75\hsize}{!}{\includegraphics{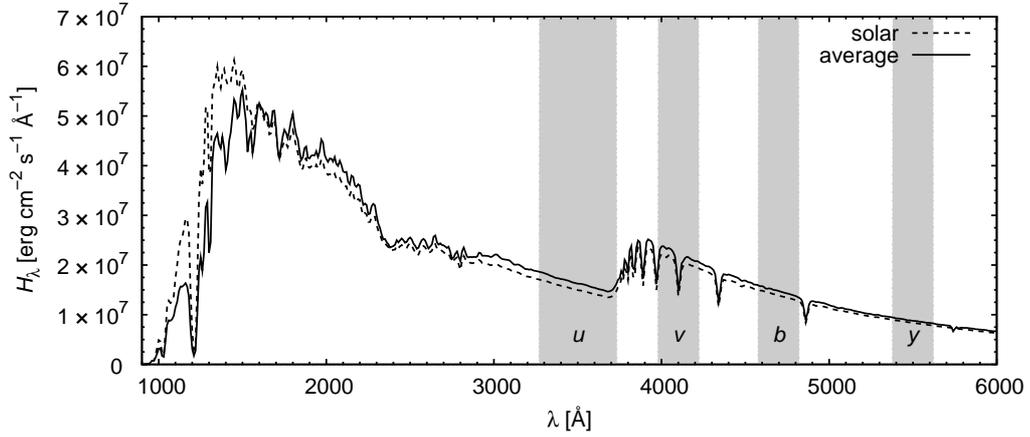}}
\caption{Comparison of the mean predicted flux (averaged over the rotation
period, solid line) and the flux roughly corresponding to the solar chemical
composition ($\varepsilon_\text{He}=-1.0$, $\varepsilon_\text{Si}=-4.25$,
$\varepsilon_\text{Cr}=-6.4$, and $\varepsilon_\text{Fe}=-4.4$). Fluxes
were smoothed with a Gaussian filter with dispersion of $10\,$\AA.}
\label{ptoksr}
\end{figure*}

As can be seen from Fig.~\ref{ptoksr}, the mean flux from \hvezda\ does not
correspond to the flux calculated for the solar chemical composition. The
redistribution of the flux from the far-UV to near-UV and visible regions is
apparent even when comparing the average fluxes. Consequently, \hvezda\ is
fainter than the normal stars with the same effective temperature in the far-UV
on wavelengths lower than $2400\,$\AA, whereas it is brighter than
normal stars in the near-UV and visible regions.

\subsection{Monochromatic variations}

\begin{figure*}[tp]
\centering \resizebox{\hsize}{!}{\includegraphics{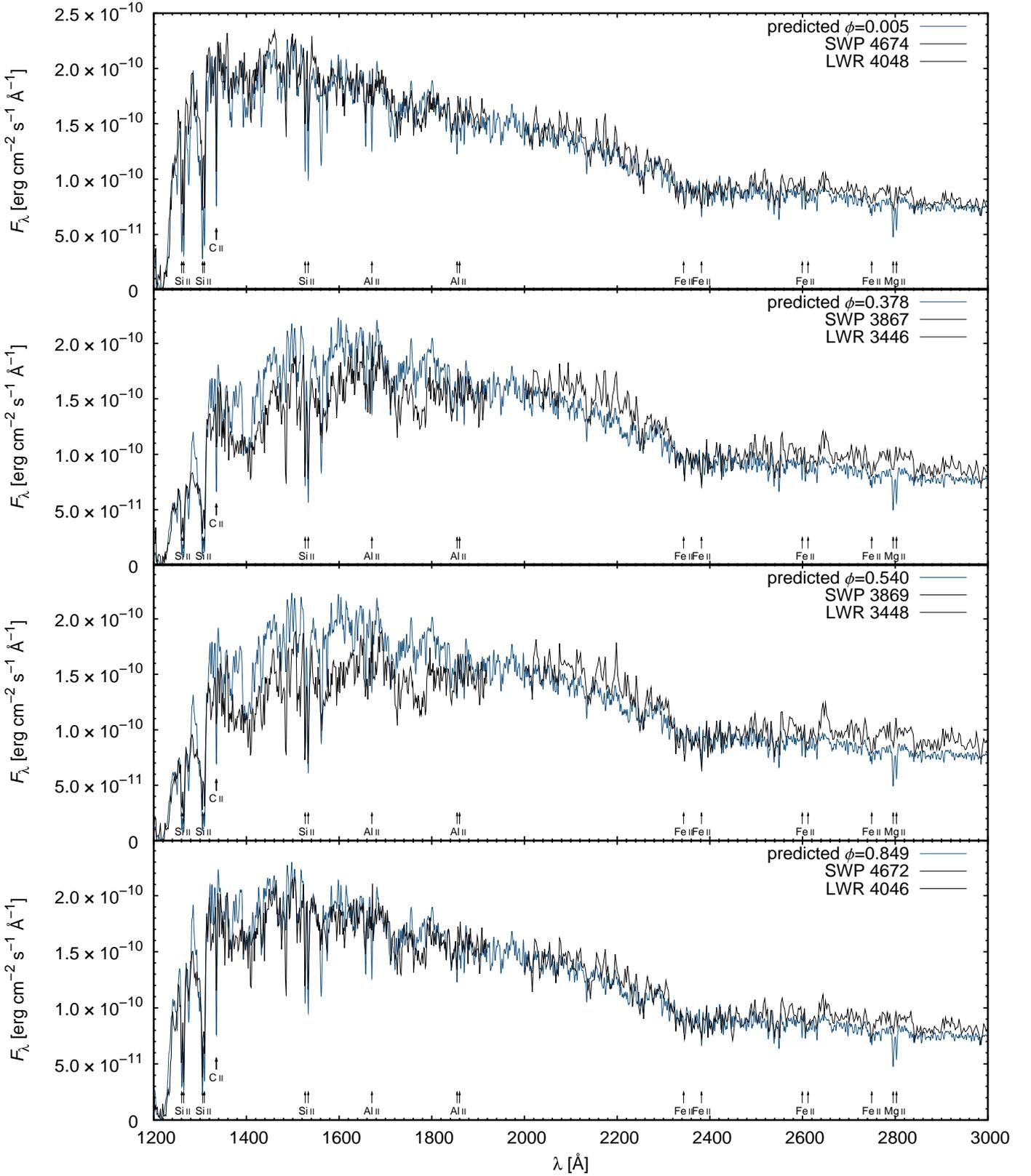}}
\caption{Predicted and observed (IUE) flux in selected phases. Individual strong
lines and iron line blends are identified.}
\label{mono}
\end{figure*}

Although the comparison of the narrow-band variations revealed the regions where
the disagreement between the observed and predicted flux variations occurs, the
narrow-band variations are inadequate to figure out the origin of these
variations. For this purpose the monochromatic variations are much more
convenient. To compare the monochromatic fluxes, we smoothed the predicted flux
variations with a Gaussian filter with a dispersion of 1.3\,\AA, which roughly
corresponds to a broadening of IUE data, and compared it with observed flux
variations. The resulting monochromatic UV fluxes are given in Fig.~\ref{mono}.

Most features that appear as individual lines in Fig.~\ref{mono} are in fact
blends of a large number of individual lines (iron is a typical case).
Only in a few cases it is possible to identify individual lines. The most numerous
individual lines are those of silicon. The monochromatic fluxes in
Fig.~\ref{mono} show that even though the continuum far-UV flux in the
region $\lambda<1500\,$\AA\ is relatively well fitted, there are some
differences in the strengths of silicon lines. That the strength of the
\ion{C}{ii} 1335\,\AA\ doublet varies in antiphase with silicon indicates that
silicon-rich regions are carbon-poor. From the weakness of  the \ion{Al}{ii}
1671\,\AA\ line we can conclude that this element is significantly underabundant
with respect to the solar value.

The predicted and observed fluxes during
the light minimum close to the phase $\phi=0$ agree very well. However, for phases
$\phi\approx0.5$ the observed and predicted fluxes disagree in some regions,
indicating that an additional opacity source is operating here, leading to
yet another redistribution of the flux from far-UV to near-UV and visible regions. From
Fig.~\ref{mono} it is also possible to identify these missing features
that are responsible for the unexplained part of the light variability. From the plot for
$\phi=0.54$ we can conclude that the missing opacity sources are located in the
wavelength interval $1350-1800\,$\AA.

\section{Detailed analysis of observed and simulated light curves}

\subsection{Description of observed light curves}

The detailed analysis of the observed light curves in the wavelength region from
1250~\AA\ to 7600~\AA\ was made on basis of all available photometric data of
\hvezda, including very precise and reliable data obtained in 1987--1997
by \citet{adel92} and \citet{pyperper}, and the data derived from IUE
spectrophotometry (see Sect.~\ref{uv}). The complete list of all photometric
observations taken in 53 photometric bands that are quite evenly distributed along the
whole spectral interval obtained in 1955--2011 is published in
\citet{mikvar}. For the calculation of photometric phases we used a new ephemeris
that takes into account the long-term variability of the period (see
Sect.~\ref{stelpa}). The quantity and the quality (the mean weighted uncertainty of
an individual photometric measurement is 5.3 mmag) of this photometric material
enable us to investigate the photometric behaviour of \hvezda\ with an
unprecedented accuracy.

First we studied the properties of the light curves in individual
passbands. Our results show that the shapes of the light curves
remained constant during the last half century. Each of them can be
well expressed by a smooth single wave curve -- a harmonic function of
a low order. Nevertheless, the shapes of the light curves in different
bands are apparently different. Applying the weighted advanced
principle component analysis (APCA) to the parameters of the harmonic
fits \citep[which allows one to find hidden relationships among them, for
details see e.g. a brief introduction in][]{mikpca} we arrived at two
principal conclusions: 1)\,Each light curve studied can be
satisfactorily well fitted by a second-order harmonic polynomial --
the amplitudes of the third and higher harmonics are always bellow 0.5
mmag. 2)\,All light curves studied can be well expressed by a linear
combination of only two basic light curves $F_1(\phi),\ F_2(\phi)$,
the amplitude of the third and higher principal components do not
exceed 1.0 mmag.

This allows us to build a relatively simple two-component
phenomenological model with a minimum of free parameters valid for all
studied light curves, where each of them is expressed by a linear
combination of two principle functions $F_1(\phi),\ F_2(\phi)$
of a rotational phase $\phi$
\begin{equation} \label{modelLC}
\Delta
m(\phi,\lambda)=A(\lambda)\hzav{\sin\zav{\frac{\pi}{2}\psi(\lambda)}
F_1(\phi)+ \cos\zav{\frac{\pi}{2}\psi(\lambda)}F_2(\phi)},
\end{equation}
where $A(\lambda)$ is the effective semiamplitude in the band centred
on the wavelength $\lambda$, $\psi(\lambda)$ is the parameter
explicitly determining the shape of the light curve in the wavelength
$\lambda$. The functions $F_1(\phi),\ F_2(\phi)$ are determined by
parameters $\gamma_1,\ \gamma_2,\ \phi_{11},\ \phi_{12},\
\phi_{21},\rm{and}\ \phi_{22}$
\begin{equation} \label{PCAF}
F_l(\phi)=\cos(\gamma_l)\,\cos[2\,\pi\,(\phi\!-\!\phi_{l1})]+\sin(\gamma_l)\,
\cos[4\,\pi\,(\phi\!-\!\phi_{l2})],
\end{equation}
where $l=$1 and 2. Functions $F_1(\phi),\ F_2(\phi)$ were determined by means of
APCA, where we confined ourselves to the first two principle components. They
represent the pair of normalised mutually orthogonal vectors in the 4-D space of
the Fourier coefficients. Eq.~\ref{modelLC} then means that the light curves
defined by their vector in the space of the Fourier coefficients should lay in
the plane determined by these two vectors of $F_1(\phi),\ F_2(\phi)$ functions.

\begin{figure}[t]
\centering \resizebox{1\hsize}{!}{\includegraphics{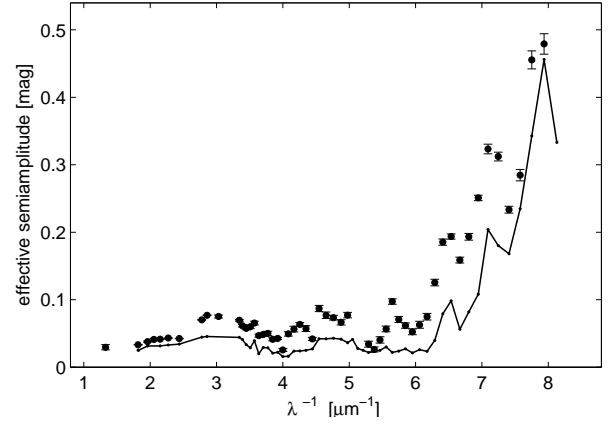}}
\caption{Comparison of the observed effective semiamplitudes
$A(\lambda)$ (see Eq. \ref{modelLC}, dots with error bars) and
semiamplitudes of simulated light variations (small dots connected by
solid line).} \label{ampl}
\end{figure}

\begin{figure}[t]
\centering \resizebox{1\hsize}{!}{\includegraphics{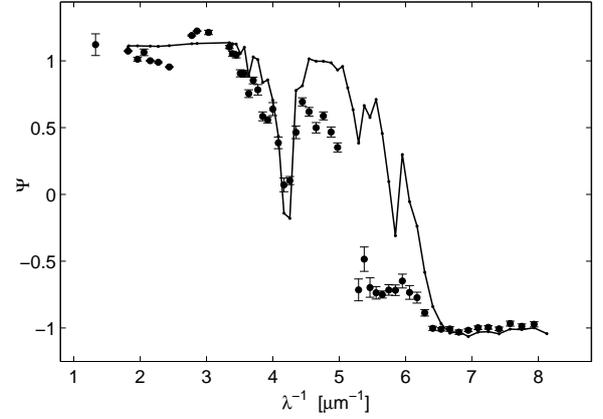}}
\caption{Comparison of the form of observed light curves described
by the dimensionless value $\psi$ (see Eq. \ref{modelLC}, dots with
error bars) and with the value predicted by our models (small dots connected by
solid line).} \label{psi}
\end{figure}

The light curves reach the maximum amplitude for wavelengths shorter
than 1600\,\AA, whereas their shapes are nearly identical in this
region (see Figs. \ref{ampl}, \ref{psi}, and Table~\ref{tabampl}).
That is why we rotated the orthogonal vector base in its plane by an
angle of $-0.12$~rad so that the $x$-axis intersects these points. The
parameters of the principle light curves $F_1(\phi),\ F_2(\phi)$ are
then $\gamma_1=0.1244$\,rad, $\phi_{11}=-0.0132,\ \phi_{12}=0.00442$,
$\gamma_2=0.2468$\,rad, $\phi_{21}=-0.2655$, and $\phi_{22}=0.0912$.

\subsection{Comparison of observed and predicted light curves}

The shapes of the simulated light curves can be described by the same procedure as the
observed ones, consequently we can compare them unambiguously. We have arrived at
the following conclusions:
\begin{enumerate}
\item The shapes of both the observed and the simulated light curves strongly
depend on the wavelength. There are differences not only in the amplitudes of
variations $A(\lambda)$ but also in their shapes, quantified by the
parameter $\psi(\lambda)$. This can be understood as the result of the
influence of several elements (at least Si, Fe, Cr).
\item The shapes of the basic phase curves $F_1(\phi),\,F_2(\phi)$
and the dependence of their parameters $A(\lambda)$ and
$\psi(\lambda)$ on wavelength can be caused by
two vast overlapping photometric spots centred at phases
$\phi_{01}=0.3209(24)$ and $\phi_{02}=0.6244(23)$ with different
contrasts on the stellar surface. These spots can be identified as
spots with overabundant Si and iron-peak elements (Cr, Fe).
\item The uneven colours of the photometric spots described by
the parameter $\psi$ are illustrated in Fig.~\ref{fuvnuv} with the
IUE light curves in wavelengths $1500\,\AA,\ 2100\,\AA$, and
$3000\,\AA$.
\item The dissimilar spectral energy distribution in spots is the
reason why we do not find any `null' point with zero amplitude
of the flux variations.
\item The simulation of the light variations cannot fully explain the observed
amplitudes in the whole studied spectral region. The simulated
amplitudes are on average smaller by a few tens of percent than the
observed ones (see Fig.\,\ref{ampl}), which implies that we have neglected
some of the important sources of the light variations in our modelling.

\end{enumerate}

\section{Discussion}
\label{kecame}

Despite the overall good agreement between the predicted and observed light
curves, there still remain some differences between the simulations and
observations. These are especially apparent in the Str\"omgren $u$ filter and in
the UV region $2000-2500\,$\AA. Note, however, that \hvezda\ has the highest
amplitude of all stars modelled in detail so far, consequently it is encouraging
that even this high amplitude can be explained  for the most part. Here we
discuss some possible reasons for the remaining disagreements.

\subsection{Limitations of abundance maps}

The limitations of abundance maps used could be an important source of
discrepancy between prediction and observation. While the fine structure of
abundance maps does not significantly influence the predicted variability (as
shown by \citealt{myhr7224} using different abundance maps of \citealt{leh2}),
other effects may be significant. The predicted SED variations are sensitive to
the maximum abundance and abundance amplitude of a given element in the map. The
observed variations of silicon equivalent widths are nicely reproduced by the
abundance maps \citep[see Fig.~1 of][]{kus}, indicating their high reliability.
This is also supported by a good agreement between predicted and observed SED
variations in the far-UV region, where the silicon dominates. On the other hand,
the predicted and observed equivalent widths of
chromium and iron in Fig.~1 of \citet{kus} disagree to some extent. Our tests showed that especially
the modified abundance of iron could help to explain some part of the discrepancy
between theory and observations.

Another source of the discrepancies might be connected with
limitations of the model atmospheres used for the abundance analysis. All
these considerations point to a need of new more precise abundance
maps of \hvezda.

\subsection{Influence of additional elements}

We have shown that most of the light variability of \hvezda\ is caused by the
uneven distribution of chemical elements. Consequently, it is possible that a
part of the remaining discrepancy between theory and observation is also caused
by other element(s), whose surface distribution was not mapped by \citet{kus}.
Both the optical and UV light curves provide strong constraints on the opacity
caused by this as yet unidentified element, which might redistribute the flux from
UV region of about $2000-2500\,$\AA, especially to the region of Str\"omgren
$u$.

We tested if either of the chemical elements currently included in the TLUSTY model
atmospheres could cause these light variations. Excluding magnesium because of its
low abundance previously, no other element included in TLUSTY
(i.e., C, N, O, Ne, Al, and S) is able to cause the remaining light variations
observed
in \hvezda. Consequently, it is likely that another element (especially the
iron-peak ones) could be the cause.

Very recent observations in a broad spectral range suggest that titanium and
oxygen could also be contributors to inhomogeneous abundance structures on the
surface of \hvezda. Especially titanium (provided it is significantly
overabundant) is one of the potential causes of the UV variations we cannot
simulate so far.

\subsection{Vertical abundance stratification}

Vertical abundance stratification is observed in some CP stars
\citep[e.g.,][]{rypop}. \citet{sokolja} proposed that the vertical abundance
stratification may influence the light variability. Our test calculations
confirmed these expectations. The models with overabundant iron in
the outer regions (for the Rosseland optical depth $\tau_\text{Ross}<0.1$)
indeed show a larger magnitude difference in $u$ than in the $v$, $b$, and $y$
colours. Consequently, vertical abundance stratification could possibly explain
the difference between the phases of maxima in individual Str\"omgren filters.
However, the influence of vertical abundance stratification on the UV region is
relatively low, consequently another process is needed to explain the
difference between observation and theory in this region.

\subsection{Surface temperature variations}

Surface temperature differences and variable temperature gradients were also
suggested as possible causes of the CP star light variability
\citep{biltep,steptep}. Hot stars may retain subsurface convection zones
\citep{kant}, which can generate local magnetic fields. These
magnetic fields may give rise to the surface temperature differences, and
consequently cause the light variability. However, mild differences between the
observed and predicted light curves do not indicate any 
(effective) temperature differences on the surface of \hvezda. This is supported
also by the fact that e. g. the predicted mean of the flux distribution 
simulates the observed one fairly well, as depicted in Fig.~\ref{ptok}. Moreover, a good
agreement between the observed and predicted light curves in the far-UV region and
the $vby$ light curves in the visible region excludes the temperature
differences as a main source of the light variability.

\subsection{Influence of the turbulent velocity}

In our study we assumed a generic value of the microturbulent velocity
$2\,\text{km}\,\text{s}^{-1}$. This parameter, which roughly
accounts for atmospheric velocity fields, likely has a zero value in
corresponding normal stars \citep{lak}. We kept a nonzero value here
as a very rough approximation of Zeeman line broadening. The 
adopted value of the microturbulent velocity may, however, influence
the emergent flux. At higher microturbulent velocities the line
transitions are able to absorb radiation more effectively, increasing
thus the temperature in the continuum forming regions. To estimate the
magnitude of this effect, we calculated an additional model with higher
microturbulent velocity $4\,\text{km}\,\text{s}^{-1}$ and assuming
enhanced abundance of heavier elements ($\varepsilon_\text{He}=-1.0$,
$\varepsilon_\text{Si}=-2.25$, $\varepsilon_\text{Cr}=-4.9$,
$\varepsilon_\text{Fe}=-3.4$). We compared the resulting flux
distribution with the model with the same chemical composition,
but with a standard microturbulent velocity of
$2\,\text{km}\,\text{s}^{-1}$. The calculated magnitude difference
(Eq.~\ref{velik}) between these models has its minimum $-0.05\,$mag in
the near-UV region $3000-3800\,$\AA\ and a maximum about $-0.10\,$mag
in the region $2200-2550\,$\AA. These are the regions where the most
apparent differences between observed and predicted light variations
occur. Consequently, a higher value of the microturbulent velocity
and/or surface microturbulent velocity distribution (see
Sect.~\ref{slunik}) cannot be ruled out as a possible cause of the
remaining difference between theory and observation.

%

\subsection{The effects of fast rotation}
\label{rot}

\hvezda\ belongs to the fast rotators among CP stars. From Table~\ref{hvezda} we
can infer its rotational velocity $v_\text{rot}=320\,\text{km}\,\text{s}^{-1}$,
indicating a rotational velocity close to the critical one. To quantify
this, the stellar radius has to be known with sufficiently high precision.
Using the stellar parameters derived from spectroscopy and
photometry (i.e., $v_\text{rot} \sin i$, inclination, and period, see
Table~\ref{hvezda}), we can estimate the equatorial radius to be
$R_\text{eq}=3.3\pm0.6\,{R}_\odot$. This value agrees well with
the stellar radius derived from the evolutionary tracks in the $T_\text{eff}-
\log g$ plane of \citet{salek}, which is $R=3.2\pm0.5\,{R}_\odot$ (the
derived mass is $M=3.8\pm0.2\,{M}_\odot$). Note, however, that this yields
a significantly higher radius than that derived from the observed UV flux, which
is $R=1.9\pm0.1\,{R}_\odot$ assuming a distance of $79\pm1\,\text{pc}$
\citep{hipik}. This possibly points either to a problem with the absolute flux
calibration, or to an inclination that is too low. The latter is supported by a lower
radius of $R=2.3\pm0.1\,{R}_\odot$ derived from photometry and evolutionary
tracks by \citet{kobacu}.

To study the effect of fast rotation at its extremum, we assumed the equatorial
radius $R_\text{eq}=3.3\,R_\odot$, and $M=3.8\,M_\odot$, the polar radius is
then $R_\text{p} =R_\text{eq} \zav{1+v_\text{rot}^2R_\text{eq}/\zav{2GM}}^{-1}
=2.7\,R_\odot$ \citep{dvojka}. The ratio of the rotational velocity to the
critical one is then $v_\text{rot}/v_\text{krit}=
v_\text{rot}/\sqrt{{2GM}/{3R_\text{p}}}= 0.75$. If the star rotates this
rapidly,
the polar to equator difference in the effective surface gravity and effective
temperature are $\Delta\log g=
\log\zav{GM/R_\text{p}^2}-\log\zav{GM/R_\text{eq}^2-v_\text{rot}^2/R_\text{eq}}
\approx-0.4$ and $\Delta T_\text{eff} \approx3000\,\text{K}$ (assuming
$T_\text{eff}(\vartheta)\sim g^{1/4}(\vartheta)$, \citealt{vonz,slat,owodisk}).

The variations of local surface gravity and effective temperature
connected with a fast rotation by itself cannot raise any light
variability (assuming fixed axis of rotation) because of their axial
symmetry. However, even axisymmetric surface variations may modify the
light variations caused by inhomogeneous elemental surface
distribution via several effects. First, surface layers with different
abundances may respond differently to the temperature and surface
gravity variations, modifying the flux distribution. Second, the area
of the surface element is modified through the oblateness of the
surface of a rotating star. Finally, the direction of the beam
pointing to the observer relative to the local outward normal and its
cross-section are different on spherical and oblate surface.

To test the influence of these effects, we calculated an additional grid of
model atmospheres corresponding to the stellar equator
($T_\text{eff,eq}=11\,500\,\text{K}$, $\log g_\text{eq}=3.71$) and to the  pole
($T_\text{eff,p}=14\,300\,\text{K}$, $\log g_\text{p}=4.16$) with different
abundances of silicon and iron (after Table~\ref{esit}). The polar and
equatorial effective temperature and gravity were calculated assuming that the
\hvezda\ parameters $T_\text{eff}$, $g$ derived by \citet{kus} represent some
kind of mean over the stellar surface and that these parameters correspond to
some particular surface region. In this case $T_\text{eff}(\vartheta)=
 \hzav{g(\vartheta)/g}^{1/4}T_\text{eff}$. For simplicity we assumed a fixed
abundance of helium and chromium here ($\varepsilon_\text{He}=-1$,
$\varepsilon_\text{Cr}=5.9$) because these elements are not the main sources of the
light variability. The resulting light curve was derived using Eq.~\ref{velik},
now interpolating the intensities in Eq.~\ref{barint} also between the models
corresponding to the pole and to the equator.

The inclusion of the gravity darkening only \citep[with the latitude-dependent
radius calculated after][]{harcog} leads to the decrease of predicted amplitude
of light variations by up to $0.01\,$mag. This is because for
both silicon and iron the values of $\Delta m_\lambda$ Eq.~\ref{velik}
decrease with decreasing temperature.

The inclusion of different surface areas caused by the oblateness of the star leads
to another small modification of the light curve, which is mostly negligible
(the difference is up to one millimagnitude). We did not consider the effect of
the difference of beams pointing to the observer (in an oblate star compared to
the spherical one), but given the small influence of other effects, we expect
that this is also negligible.

We conclude that the effect of the gravity darkening modifies the light curves,
but it does not seem to be the main reason for the discrepancy between
predicted and observed light curves. But the effect of gravity darkening on the
light curves introduces a new possibility to test the theory of gravity darkening
in the future. To this end, the influence of this effect on the abundances
derived from the Doppler imaging has also to be accounted for.

\subsection{Influence of the magnetic field}

\hvezda\ is known to host a large-scale magnetic field, which could, in
principle, contribute to the light variability via the rotational modulation of
the magnetic field intensity and thus opacity in Zeeman broadened spectral
features. Assuming a dipolar field geometry \citet{trigilio} based on
phase-resolved longitudinal magnetic field measurements from
\citet{borra-landstreet-1980} determined a polar magnetic field $B_{\rm p}=3$~kG
and the angle between the stellar rotation axis and the line-of-sight
$\beta=74\degr$. Regarding magnetic CP stars, this is a fairly moderate magnetic
field and, as demonstrated by \citet{zeeman-paper2} based on detailed model
atmosphere calculations with anomalous Zeeman effect and polarized radiative
transfer, such a field does not produce significant effects on parameters
observed in different photometric systems. In addition, with the inclination
angle $i=30\degr$ the surface average magnetic field modulus varies only by
$\Delta\,\langle\,|\boldsymbol{B}|\,\rangle\approx300$~G during the rotational
cycle, which is far too small to induce any detectable changes in light-curves.

On the other hand, as shown by \citet{shulyak-2008}, the combined impact of the
magnetic field and the realistic chemistry could be more important than the
effect of using only individual abundances. Thus, the effect of including the
magnetic field in the computations of photometric parameters would be more
stronger in regions of spots with enhanced abundances compared to the case of a
homogeneously stratified atmosphere.

Because we are mostly interested in the estimate of the maximum possible
amplitude that the magnetic field could introduce in the light curves relative
to the inhomogeneously distributed abundances, we computed several magnetic
model atmospheres using the \textsc{LLmodels} code \citep{llm}. Anomalous Zeeman
splitting and polarised radiative transfer were included as described in
\citet{zeeman-paper2}. We found that the highest impact of the magnetic field
for a spot with enhanced Fe, Si, and Cr, located at the pole with $|\boldsymbol{B}|=3$~kG and equator with $|\boldsymbol{B}|=1.5$~kG amounts to
$\Delta u\approx0.005$~mag. This difference subsequently decreases for other
Str\"omgren filters. Accounting for the surface averaged magnetic field modulus
results in $\Delta\,\langle\,|\boldsymbol{B}|\,\rangle=2.2$~kG at the phase when
the magnetic pole is visible and $\Delta\,\langle\,|\boldsymbol{B}|\,\rangle=1.9$~kG
at the phase of magnetic equator respectively. The effect is then reduced to
$\Delta u\approx0.001$~mag. All this is of an order of magnitude less than
needed to explain the deviations between observed and predicted non-magnetic
amplitudes of the $u$-parameter shown in Fig.~\ref{cuvir_hvvel}. We therefore
conclude that the magnetic field has little or negligible effect on the
light-curve appearance and can be ignored in the present study.

\subsection{Convection zone in helium-rich models}
\label{slunik}

In addition to the central convection zone, a hot star may have subsurface iron and
helium convective zones \citep{maedsam,magem,kant}. Our models show that for
helium-rich models with $\varepsilon_\text{He}\gtrsim0$, the helium convection zone
moves towards the stellar surface and its top is located in the model atmosphere
for Rosseland optical depths $\tau_\text{Ross}\approx1-100$. The existence of
a subsurface convection zone may have interesting astrophysical consequences
\citep[e.g.,][]{magem,kant}. The convective zone may create a dynamo, generating
the chromospheric activity and consequently also X-ray emission, whose existence
is still puzzling in A type stars \citep{rentgena}. Moreover, convection may
cause surface turbulence, leading to inhomogeneous distribution of the turbulent
velocity provided helium is also distributed inhomogeneously.

However, these effects are likely strongly damped in \hvezda, which has a strong
surface magnetic field \citep{laborka}. Our results show that the magnetic field
energy density dominates over the gas energy density to the optical depths of
about $\tau_\text{Ross}\approx10^3$, suppressing any motion perpendicular to the
magnetic field lines.

\subsection{NLTE effects}

Although the TLUSTY model atmosphere code enables us to calculate NLTE models,
we confined ourselves to LTE models, because we expect NLTE effects to be
marginal for the light variability. To test this we calculated additional NLTE
models. Because we do not have sufficient atomic data to calculate NLTE models
including chromium, we forced LTE for this element, even if for the remaining
elements we assumed NLTE.

We calculated an NLTE model with enhanced abundance of heavier elements
($\varepsilon_\text{He}=-1.0$, $\varepsilon_\text{Si}=-2.25$,
$\varepsilon_\text{Cr}=-4.9$, $\varepsilon_\text{Fe}=-3.4$) and compared it with the
corresponding LTE model. The resulting fluxes differ by about $1-2\,\%$ in the
visible and near-UV regions. The most pronounced changes appear in the far-UV region
with $\lambda<1500\,$\AA. Consequently, the changes owing to NLTE are
significantly lower than those owing to variable abundances, and cannot be the
main source of the difference between theory and observations.

The NLTE effects are, however, significant in lines. The NLTE effects lead not
only to the well-known strengthening of the core of hydrogen lines
\citep[e.g.,][]{bstar2006}, but the line equivalent widths of other elements are
also slightly affected. Finally, in NLTE some infrared lines appear in emission.

\section{Conclusions}

We successfully simulated the UV and visual SED variability of the helium-weak star
\hvezda. We assumed that the light variability is caused by the inhomogeneous
surface distribution of elements and used model atmospheres to predict the light
variability.

Individual chemical elements are distributed inhomogeneously on the surface of
\hvezda, as derived by \citet{kus} by Doppler mapping. The chemical composition
influences the emergent flux through flux redistribution from the far-UV to
near-UV and visible regions.  The bound-free transitions of silicon and the
bound-bound transitions of iron and chromium mostly cause the flux
redistribution in \hvezda. As a result of the flux redistribution,
the individual surface elements display different brightness in individual
photometric bands, although the total (frequency integrated) emergent flux is
the same for all surface elements. The inhomogeneous surface brightness
manifests itself by the light variations caused by stellar rotation.

The inhomogeneous surface distribution of silicon, chromium, and iron is able to
explain most of the observed UV and visible SED variations. We successfully
reproduced the antiphase behaviour of the light curves in the far-UV and visible
regions. We emphasise that the variability seen in the visible is just a faint
gleam of the variability seen in the UV. While the amplitude of the light
curves merely reaches about a few hundredths of magnitude in the visual domain, it
reaches about 1\,mag in the UV. However, our models are able to reproduce just
part of the variability seen in the UV region $2000-2500\,$\AA\ by IUE and in
the $u$ filter of the Str\"omgren photometric system. Another mechanism(s) has
to be invoked to explain this difference between the observed and predicted
light curves. This so far unidentified mechanism could be connected with
inhomogeneous distribution of an element whose surface abundance distribution
was not mapped by \citet{kus}, but also other effects may contribute to this
difference.

Our models nicely reproduce the observed SED for the phase $\phi\approx0$,
when the regions with lowest elemental abundances are seen. On the other hand,
there are some discrepancies between simulated and observed SED  for the phase
$\phi\approx0.5$, when the overabundant regions appear. These discrepancies have
likely the same origin as the differences of the light curves.

Our study provides additional evidence that the light variability of
chemically peculiar stars is mostly caused by the inhomogeneous surface
distribution of individual chemical elements, flux redistribution, and stellar
rotation. The comparison of observed and predicted SED and its variation may
serve as a test of opacity sources included in current model atmospheres.
Finally, as a byproduct, it provides an image of the stellar surface, which
(besides the interferometry) is the only way how these images can be derived.

\begin{acknowledgements}
We wish to thank Dr.~P.~\v Skoda for his continuous and enthusiastic
work on the Virtual Observatory project. This work was supported by
grants GAAV IAA301630901, MEB 061014/WTZ CZ 10-2010, Deutsche
Forschungsgemeinschaft (DFG) Research Grant RE1664/7-1 to DS, and VEGA
2/0074/09. This research was partly based on the IUE data derived from
the INES database using the SPLAT package. This research made use of
NASA's Astrophysics Data System, the SIMBAD database, operated at the
CDS, Strasbourg, France and the on-line database of photometric
observations of mCP stars \citep{mikdata}. The access to the
METACentrum (super)computing facilities provided under the research
intent MSM6383917201 is also acknowledged.
\end{acknowledgements}

\clearpage
\appendix

\section{Long tables}

\begin{table}[h]
\caption{List of the IUE observations of \hvezda}
\label{iuetab}
\begin{center}
\begin{tabular}{cccc}
\hline
Camera & Image & Julian date &  Phase\\
        &&   2,400,000+\\
\hline
LWR & 3444 & 43883.94545 & 0.207\\
LWR & 3445 & 43883.98646 & 0.286\\
LWR & 3446 & 43884.03179 & 0.373\\
LWR & 3447 & 43884.07508 & 0.456\\
LWR & 3448 & 43884.11588 & 0.535\\
LWR & 4044 & 43949.80082 & 0.687\\
LWR & 4045 & 43949.84187 & 0.766\\
LWR & 4046 & 43949.88284 & 0.844\\
LWR & 4047 & 43949.92253 & 0.921\\
LWR & 4048 & 43949.96366 & 1.000\\
SWP & 3864 & 43883.91128 & 0.142\\
SWP & 3865 & 43883.95163 & 0.219\\
SWP & 3866 & 43883.99343 & 0.300\\
SWP & 3867 & 43884.03658 & 0.382\\
SWP & 3868 & 43884.07952 & 0.465\\
SWP & 3869 & 43884.12081 & 0.544\\
SWP & 4670 & 43949.80501 & 0.695\\
SWP & 4671 & 43949.84709 & 0.776\\
SWP & 4672 & 43949.88714 & 0.853\\
SWP & 4673 & 43949.92677 & 0.929\\
SWP & 4674 & 43949.96918 & 0.010\\
\hline
\end{tabular}
\end{center}
\end{table}

\begin{table}[h]
\caption{%
Observed and simulated effective semiamplitudes of light curves and
parameters describing their forms in individual photometric bands of ultraviolet
and optical regions (see Eq.\,\ref{modelLC}).}
\begin{center}
\begin{tabular}{clrrrrr}
  \hline
  band &  $\lambda$ & \multicolumn{3}{c}{Observation} &
  \multicolumn{2}{c}{Simulation} \\
 &  & $N\ $ & $A(\lambda)$ \ &
 $\psi(\lambda)\quad$& $A(\lambda)$ & $\psi(\lambda)$ \\
 & [\AA] & &[mmag]&&[mmag]\\
  \hline
 &1230&    &      &          &333&$-1.04$\\
&1260&  11&479(15)&$-0.97(2)$&456&$-1.00$\\
&1290&  11&455(14)&$-0.99(2)$&343&$-1.01$\\
&1320&  11& 285(9)&$-0.97(2)$&235&$-1.01$\\
&1350&  11& 233(5)&$-1.01(2)$&168&$-1.04$\\
&1380&  11& 312(7)&$-0.99(2)$&180&$-1.03$\\
&1410&  11& 323(7)&$-1.00(2)$&204&$-1.03$\\
&1440&  11& 251(4)&$-1.02(1)$&108&$-1.06$\\
&1470&  11& 193(5)&$-1.03(2)$& 82&$-1.03$\\
&1500&  11& 159(5)&$-1.01(2)$& 56&$-1.04$\\
&1530&  11& 194(4)&$-1.01(2)$& 98&$-0.97$\\
&1560&  11& 185(5)&$-1.00(2)$& 79&$-0.84$\\
&1590&  11& 125(5)&$-0.89(3)$& 40&$-0.58$\\
&1620&  11&  75(5)&$-0.77(4)$& 24&$-0.24$\\
&1650&  11&  63(5)&$-0.73(5)$& 26&$-0.05$\\
&1680&  11&  52(5)&$-0.65(5)$& 21&$0.30 $\\
&1710&  11&  62(4)&$-0.72(4)$& 27&$-0.31$\\
&1740&  11&  71(5)&$-0.72(4)$& 24&$0.10 $\\
&1770&  11&  97(4)&$-0.75(3)$& 22&$0.46 $\\
&1800&  11&  57(5)&$-0.74(5)$& 30&$0.71 $\\
&1830&  11&  40(5)&$-0.70(8)$& 25&$0.58 $\\
&1860&  11&  26(4)&$-0.48(9)$& 23&$0.66 $\\
&1890&  11&  34(5)&$-0.71(8)$& 22&$0.38 $\\
\hline
&1920&  &  & & 25&0.63 \\
&1950&  &  & & 28&0.80 \\
&1980&  &  & & 41&0.96 \\
&2010&  10&  77(4)& 0.35(4)& 36&0.93 \\
&2050&  10&  66(4)& 0.47(4)& 41&0.96 \\
&2100&  10&  73(4)& 0.59(3)& 43&1.00 \\
&2150&  10&  77(5)& 0.50(4)& 42&1.00 \\
&2200&  10&  87(5)& 0.62(4)& 42&1.01  \\
&2250&  28&  42(3)& 0.69(3)& 27& 0.81 \\
&2300&  10&  57(5)& 0.46(5)& 25& 0.78 \\
&2350&  10&  63(3)& 0.10(3)& 24&$-0.18$ \\
&2400&  10&  56(5)& 0.07(5)& 24&$-0.14$ \\
&2450&  10&  49(4)& 0.39(5)& 16& 0.44 \\
&2500&  28&  25(3)& 0.64(5)& 16& 0.70 \\
&2550&  10&  43(2)& 0.56(3)& 22& 0.86 \\
&2600&  10&  41(3)& 0.58(4)& 21& 0.84 \\
&2650&  10&  50(4)& 0.78(4)& 29& 1.01 \\
&2700&  10&  48(2)& 0.85(3)& 29& 1.03 \\
&2750&  10&  47(3)& 0.75(3)& 20& 0.89 \\
&2800&  10&  65(3)& 0.90(3)& 39& 1.10   \\
&2850&  10&  60(3)& 0.91(3)& 29& 1.05   \\
&2900&  10&  58(2)& 1.04(2)& 33& 1.12   \\
&2950&  10&  61(2)& 1.05(2)& 41& 1.13   \\
&2990&  10&  69(2)& 1.10(1)& 44& 1.14   \\
&3300&  38&  75(3)& 1.21(2)&   &         \\
\hline
$u$&3500& 986&76.9(3)& 1.22(1)&45& 1.13\\
$U$&3600& 738&70.1(5)& 1.19(1)&45& 1.13\\
$v$&4100&1130&42.4(2)& 0.95(1)&34& 1.11\\
$B$&4400& 992&43.1(3)& 0.99(1)&33& 1.11\\
$b$&4650&1059&41.5(2)& 1.00(1)&31& 1.11\\
$\beta$&4860&  95&  41(2)& 1.06(3)&  \\
$\it{Hp}$&5100& 112&37.8(9)& 1.01(2)&31& 1.11\\
$y+V$&5500&2251&33.2(2)& 1.07(1)&25& 1.11    \\
$R$&7530&  32&  30(4)& 1.12(8)&  \\
\hline \label{tabampl}
\end{tabular}
\tablefoot{$N$ denotes number of observations}
\end{center}
\end{table}

\end{document}